\documentclass[conference]{IEEEtran}
\makeatletter
\def\ps@headings{%
\def\@oddhead{\mbox{}\scriptsize\rightmark \hfil \thepage}%
\def\@evenhead{\scriptsize\thepage \hfil \leftmark\mbox{}}%
\def\@oddfoot{}%
\def\@evenfoot{}}
\makeatother
\IEEEoverridecommandlockouts

\usepackage{geometry}
\usepackage{cite}
\usepackage{amsmath,amssymb,amsfonts}
\usepackage{algorithmic}
\usepackage{graphicx}
\usepackage{textcomp}
\usepackage{xcolor}
\usepackage{comment}
\usepackage{tabularx}

\usepackage{caption}
\usepackage{subcaption}
\usepackage{multirow}
\usepackage{multicol}
\usepackage{array}
\usepackage{float} 
\usepackage{adjustbox}

\def\BibTeX{{\rm B\kern-.05em{\sc i\kern-.025em b}\kern-.08em
    T\kern-.1667em\lower.7ex\hbox{E}\kern-.125emX}}

\makeatletter 
\newcommand{\linebreakand}{%
  \end{@IEEEauthorhalign}
  \hfill\mbox{}\par
  \mbox{}\hfill\begin{@IEEEauthorhalign}
}
\makeatother 

\begin{document}

\title{
Enhancing ML-Based DoS Attack Detection Through Combinatorial Fusion Analysis

}

\author{\IEEEauthorblockN{Evans Owusu}
\IEEEauthorblockA{
\textit{Fordham Univ, NY USA}\\
eowusu3@fordham.edu}
\and
\IEEEauthorblockN{Mohamed Rahouti}
\IEEEauthorblockA{
\textit{Fordham Univ., NY USA}\\
mrahouti@fordham.edu}
\and
\IEEEauthorblockN{D. Frank Hsu}
\IEEEauthorblockA{
\textit{Fordham Univ., NY USA}\\
hsu@fordham.edu}

\and


\IEEEauthorblockN{Kaiqi Xiong}
\IEEEauthorblockA{
\textit{Cyber Florida, FL USA}\\
xiongk@usf.edu}
\and
\IEEEauthorblockN{Yufeng Xin}
\IEEEauthorblockA{
\textit{RENCI, NC USA}\\
yxin@renci.org}
}

\maketitle

\begin{abstract}
Mitigating Denial-of-Service (DoS) attacks is vital for online service security and availability. While machine learning (ML) models are used for DoS attack detection, new strategies are needed to enhance their performance. We suggest an innovative method, combinatorial fusion, which combines multiple ML models using advanced algorithms. This includes score and rank combinations, weighted techniques, and diversity strength of scoring systems. Through rigorous evaluations, we demonstrate the effectiveness of this fusion approach, considering metrics like precision, recall, and F1-score. We address the challenge of low-profiled attack classification by fusing models to create a comprehensive solution. Our findings emphasize the potential of this approach to improve DoS attack detection and contribute to stronger defense mechanisms.
\end{abstract}


\section{Introduction} 

DoS attacks threaten online service availability and stability, making server protection crucial \cite{gupta2021distributed}. With evolving digital threats, more sophisticated and frequent attacks require strong countermeasures to secure networked infrastructure \cite{rahouti2021synguard}. While many ML models have been used, there's a need for innovative approaches to enhance detection and performance. 

While ML has shown promise in various cybersecurity applications, including denial DoS detection, it is important to recognize its limitations \cite{mittal2022deep}. Among these limitations, with regard to DoS detection, is the interpretability and explainability. Specifically, some ML algorithms, especially deep learning models, are considered black boxes, making it difficult to interpret their decision-making process. This lack of interpretability and explainability can hinder trust and make it challenging to understand why a particular decision or prediction (e.g., whether a networking flow is malicious or benign) was made.

To overcome such limitations, a holistic approach combining ML with other techniques, network-level mitigations, anomaly detection, and expert knowledge can be employed to enhance the accuracy and robustness of DoS detection systems. In this paper, we introduce a novel methodology based on leveraging the cutting-edge approach of combinatorial fusion, which harnesses recently developed algorithms and techniques for model fusion. Our objective is to combine multiple ML models using combinatorial fusion analysis to achieve superior DoS attack detection performance and interpretability.

The key contributions of this paper are summarized as follows.
\begin{itemize}
    \item 
    Examine the key ML models commonly used for DoS attack detection. Such an examination serves as the ground truth for integrating the combinatorial fusion analysis (CFA) approach into our ML-based DoS detection.
    \item CFA approach: Develop an innovative approach based on CFA for combining multiple ML models in DoS attack detection. 
    Our methodology encompasses various techniques, including advanced score combination, rank combination, weighted combination, and the consideration of diversity strength across multiple scoring systems.
    \item Conduct performance evaluations to assess the effectiveness of our CFA. Through comparative analyses, we showcase the significant improvements achieved in terms of detection accuracy, false positive rates, and overall precision.
\end{itemize}

By leveraging the power of the recently developed combinatorial fusion approach and its associated algorithms, our study aims to push the boundaries of DoS attack detection. We anticipate that our findings will contribute to the advancement of more robust and effective defense mechanisms against DoS attacks, bolstering the uninterrupted availability and security of critical online services.


 \label{sec:introduction}

\section{Related Work} 

DoS attack detection continuously gains attention in cybersecurity, with diverse methods proposed to enhance accuracy \cite{zhijun2020low}. Escalating DoS attacks prompted increased focus by researchers \cite{david2021discriminating}. 
Many studies employ ML algorithms for DoS attack detection, using models like SVMs, Random Forests, and Naive Bayes. These models analyze network patterns, using features like packet rates, sizes, and volume to differentiate normal and attack traffic \cite{ali2023machine}. Further, CNNs, RNNs, and variants extract intricate patterns and temporal dependencies from network traffic \cite{mittal2022deep}, showing promise in accurate attack identification. Ensemble methods like Bagging, Boosting, and Stacking also gained traction, combining models to enhance prediction \cite{deepa2019design}. These approaches counter individual model limitations, boosting accuracy and robustness.




In recent years, model fusion techniques, such as combinatorial fusion analysis (CFA), have emerged as practical means of combining multiple ML models' performance \cite{hsu2006combinatorial, hsu2010rank}. These approaches leverage algorithms and metrics specifically designed to integrate the outputs of different models, considering their diversity and individual performance. The goal is to leverage diverse models' collective knowledge and expertise to achieve higher accuracy and detection rates \cite{hurley2020multi}. While previous studies have remarkably contributed to DoS attack detection, our work focuses on the emerging CFA. To the best of our knowledge, this is the first work that adopts CFA for examining the performance of ML models-based DoS attack detection. By combining the strengths of multiple ML models using advanced fusion techniques, we aim to achieve more accurate and reliable detection results. \label{sec:related}


\section{Methodology} \label{sec:methodology} \subsection{Problem Abstraction}

\begin{table}[h!]
\caption{Types of network traffic in the LYCOS-IDS dataset and their training and testing sets.}
\label{tab:lycos_train_test_split}
\centering
\begin{tabular}{|lcll|}
\hline
Traffic & Encoding & Train set & Test set \\
\hline
Benign & 0 & 330474 & 110158 \\
Bot & 1 & 550 & 183 \\
DDoS & 2 & 71761 & 23920 \\
DoS Goldeneye & 3 & 5073 & 1691 \\
DoS Hulk & 4 & 119241 & 39747 \\
DoS Slowhttptest & 5 & 3649 & 1216 \\
DoS Slowloris & 6 & 4255 & 1418 \\
FTP Patator & 7 & 3001 & 1000 \\
Heartbleed & 8 & 7 & 2 \\
Portscan & 9 & 119197 & 39732 \\
SSH Patator & 10 & 2218 & 739 \\
Webattack Bruteforce & 11 & 1020 & 340 \\
Webattack Sql Injection & 12 & 9 & 3 \\
Webattack XSS & 13 & 489 & 163 \\
\hline
Total & & 660944 &  220,312 \\
\hline
\end{tabular}
\end{table}

Modern DoS attacks (e.g., UDP, DNS, SYN, NTP) jeopardize networking environments. Existing approaches target DoS detection but struggle with evolving DoS attacks. Addressing this demands advanced classification using fusion, stats, and recent attack data. This tailors defense systems to distinct attack features. Intrusion detection algorithms should evolve beyond old datasets and traditional methods. Dynamic DoS nature mandates a comprehensive approach, moving from detection to classification and effective responses \cite{tian2021lightweight}.

CFA is an emerging approach for combining results from multiple scoring systems. These systems are described by a set of statistical attributes or variables at the data level or by a group of algorithms or models at the computational informatics level. In this paper, we leverage the CFA approach to combine the results from six base ML models, namely Linear Discriminant Analysis (A), Gaussian Naive Bayes (B), Logistic Regression (C), K Nearest Neighbor (D), Decision Trees (E), and Random Forest (F). The CFA approach is a process of ensemble reinforcement learning, where all possible rank and score combinations are considered to find the optimal combination of the candidate models. We apply this approach to address the problem of DoS detection. To determine the most effective model fusion performance, our methodology utilizes the emerging CFA approaches, which involve considering both the score and rank functions for each model (i.e., scoring system) and applying metrics such as average score combination, average rank combination, and weighted combination that factor in the diversity strength or the performance of each scoring system. Last, we employ a 2-model combination to fuse 6 models, pairing them 2 at a time.

\subsection{Dataset}

We use the LYCOS-IDS2017 dataset created through the LycoSTand flow extractor \cite{rosay2021cic}. These datasets contain five days' worth of network flow entries, each comprising 83 features. In total, there are 1,837,498 entries in the dataset. To prepare the data for analysis, we used a 75\% split for the training set and 25\% for the test set as shown in Table \ref{tab:lycos_train_test_split}. The types of network traffic in the dataset and their training and testing sets are summarized by Table \ref{tab:lycos_train_test_split}. After training the six ML models on the training data, we assessed their performance on unseen data by applying the test dataset to each model. The prediction probabilities for each model across the different classes were collected. By convention, the class with the highest probability was selected as the final prediction for each data entry. These probabilities represent the models' confidence level in their predictions for the individual data items. The probabilities generated by the six scoring systems are considered scores for each individual data entry. This score information formed a new dataset, including the original data items and the six probabilities obtained from the models. The CFA algorithm leveraged this new dataset to derive valuable insights using various CFA approaches discussed later in the paper.

\subsection{Data Pre-processing and Exploratory Analysis (Probabilities and Confidence Scores)}

Probabilities and confidence scores are vital in ML models, especially for classification. They offer insight into prediction likelihood and confidence for each label. Models like logistic regression, SVM with probability outputs, and ensembles like Random Forests and Gradient Boosting provide more than just class labels. They yield probabilities or confidence scores for each class, enhancing classification information. Further, probabilities reflect a model's certainty in class predictions. A 0.8 probability for Class A indicates strong belief, while 0.2 suggests lower confidence. This nuanced information aids decisions, especially in uncertain or multi-class scenarios. Evenly distributed probabilities signal ambiguity, while high single-class probabilities show confidence.

Moreover, these probabilities assist in post-processing tasks such as thresholding, ranking, and model fusion. Thresholds adjust predictions for precision-recall trade-offs. Ranking classes by probabilities helps select top-k likely classes, important for uncertainty. Combining models' predictions via weighted voting or using probabilities as weights enhances ensemble accuracy and robustness. In this paper, we collect the probabilities associated with each prediction, the highest probability among all the classes for the various data points. Treating these probabilities as scores for each data point, we explore different CFA metrics to combine them. Our goal is to identify the optimal approach for combining these scores across multiple models to develop an effective combined model.

\subsection{Performance Evaluation of Scoring Systems}

By employing conventional ensemble methods, we can combine multiple ML classification models and evaluate their performance using specific metrics. The approaches used here are voting-based ensembles and collective confidence.

In voting-based ensembles, multiple classification models are trained independently on the same dataset. During the prediction phase, each model generates its own set of predictions for a given input, and the final prediction is determined based on a voting mechanism. There are two main types of voting-based ensembles, majority voting (each model in the ensemble casts a vote for the predicted class) and weighted voting (each model's prediction is weighted based on its performance). In collective confidence approach, instead of treating predictions as discrete class labels, we can consider each model's confidence scores. One common approach is to calculate the average probability for each class across all models and select the class with the highest average probability as the final prediction. This approach accounts for the collective confidence of the models. Alternatively, the probabilities can be weighted by the performance of each model, similar to the weighted voting approach, to assign more importance to models with better performance.

Choosing between weighted voting and using probabilities depends on the problem and model traits. Finding the right approach might need trial and error. In our paper, we employ the weighted combination of probabilities/confidence scores as the CFA metric, named weighted combination by performance, for two-model fusion. Model weights, assigned by recall scores, prioritize accurate detection of all attack instances. Lastly, Stacking is a advanced ensemble method where a meta-model learns to combine predictions from base models. Ensemble techniques enhance performance by leveraging model strengths, capturing diverse viewpoints and complex patterns. 
Careful experimentation and validation are essential to find the practical model combination for a classification problem. Next, details are provided for the key components leveraged from CFA, namely, rank score characterization (RSC) for the ML models, diversity between RSC functions (scoring systems), and rank combination vs. score combination.


A scoring system $A$ on the data set $D = \{d_1, d_2, ...,d_n\}$, comprising a score function $s_A$ and a derived rank function $r_A$, was proposed in \cite{hsu2002methods}. By sorting the values in descending order in the score function $s_A: D \mapsto \mathbb{R}$, a rank function $r_A: D \mapsto \mathbb{N}$, where $\mathbb{N}=\{1, 2, 3, ...n\}$, is obtained. The RSC function $f_A: \mathbb{N} \mapsto \mathbb{R}$ upon scoring system $A$ is expressed as follows:
\begin{equation}\label{eq:RSC}
f_A(i) = s_A(r_A^{-1}(i)) = (s_A \circ r_A^{-1})(i)
\end{equation}


\begin{figure}[h!]
	\centering
	\includegraphics[height=6.6cm, width=8.8cm]{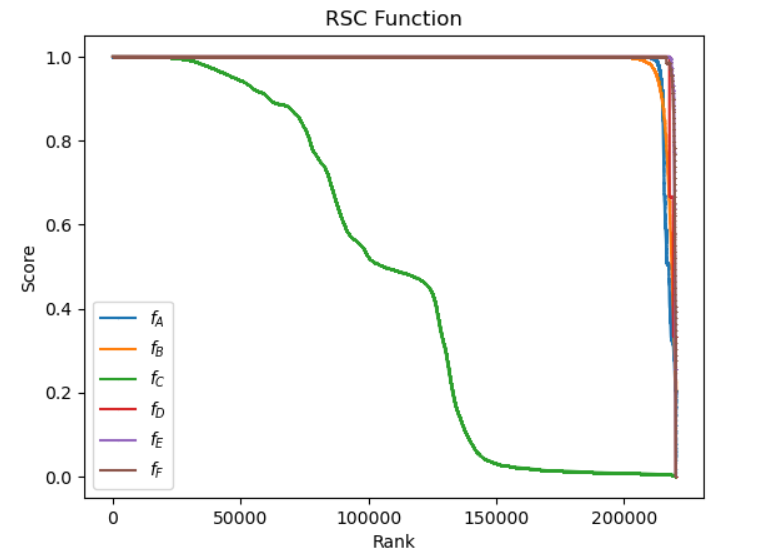}
	\caption{Rank score function graph for the six scoring systems. The area between any two RSC functions represents their diversity.}
	\label{fig:rsc_div_plot}
\end{figure}

Cognitive Diversity (CD) between the six systems (${A,B,C,D,E,F}$) is defined as the difference between the RSC functions of these systems \cite{hsu2019cognitive}. We explore the diversity of the RSC functions for each pair of these six scoring systems. Precisely, we will calculate CD between each pair of the six models. The CD between two scoring systems $A_i$ and $A_j$ is denoted by $CD(A_i,A_j)$, which is based on the RSC function of $A_i$ and $A_j$ that are denoted by $f_{A_i}$ and $f_{A_j}$, respectively. Given the rank $k \in \{1, 2, ..., n\}$, CD is defined as:

\begin{equation}
CD(A_i,A_j) = \sqrt{ \frac{1}{n^2-n} \sum_{k=1}^n{(f_{A_i}(k) - f_{A_j}(k))^2}}
\end{equation}

The diversity strength of scoring system $A$ is defined as the average CD between $A$ and all other systems. Let $D_j = \{ d_1, d_2,...,d_n\} \subseteq D$ be the labels of each cross-validation split for the scores $j \epsilon 1, ..., P$ generated by the systems, we obtain the RSC functions of our six scoring systems as: 
\begin{enumerate}
    \item The score function, $s_{kj}(d)$, gives a real number to each $d$ in $D_j$, which is the score given by the model $M_k$ to the label for $j_{th}$ split. Having the scores given by each model for each $d$ in $D_j$ provides the score function. 
    \item Sort $s_{kj}(d)$ into descending order and assigning ranks to each candidate in $D_j$ based on the sorted scores results in a rank function $r_{kj}(d)$. We rank the scores for each model/scoring system to obtain the rank function. 
    \item Compare score functions from multiple scoring systems by applying linear normalization, which is the following transformation from $s_{kj}(d):D \rightarrow R$ to $s^*_{kj}(d):D\rightarrow[0,1]$ where $s^*_{kj}(d)=\frac{s_{kj}(d) - s_{min}}{S_{max}-S_{min}}, d \epsilon D$ and $s_{max}=max \{s_{kj}(d)|d \epsilon D\}$ \\ and $s_{min}=min \{s_{kj}(d)|d \epsilon D\}$.
    \item Derive the RSC functions by sorting the normalized scores for each scoring system in descending order, using the rank values as keys (aka \textit{computational derivation of RSC function}).
    \item Plot RSC functions on the same x-y coordinate plane to depict their diversity. The x-axis and the y-axis represent the ranks and normalized scores, respectively (Fig. \ref{fig:rsc_div_plot}).
\end{enumerate}


Next, we investigate the superiority between rank combination and score combination. The rank combination can outperform the score combination under rigorous constraints/conditions \cite{hsu2010rank}. We explore the conditions for which efficient performance combinations can be obtained in favor of larger CD values between each pair of the six ML models. Here, we integrate the results of $m$ scoring systems, each with its own score function $s_{kj}(d)$ and rank function $r_{kj}(d)$ for data label $j$, where $k$ represents the scoring system index. Such techniques include score combination (SC), rank combination (RC), voting (V), average combination (AC), and weighted combination (WC). We compute SC, RC, AV, and WC based on the following weighting metrics.
\begin{enumerate}
    \item Average Combination (AC): The average score combination and average rank combination are computed as $s_s(d) = \sum_{i=1}^m [w_is_{ij}(d)]$, and $s_R(d) = \sum_{i=1}^m [w_ir_{ij}(d)]$ where $w_i = \frac{1}{m}$, and $s_s$ and $s_R$ are the score and rank functions of SC and RC respectively.
    \item Weighted combination by diversity strength (WCDS): Weighted score combination by diversity strength (WSCDS) and weighted rank combination by diversity strength (WRCDS) are the two metrics considered here, and the weights are calculated as follows.
\end{enumerate}

\begin{equation}
    W_i = \frac{\textnormal{weight of model  i}}{\textnormal{sum of weights}}  = \begin{cases}
                    \frac{1}{N}, AC \\
                    \frac{ds(A_i)}{\sum_{i=1}^N ds(A_i)}, WCDS
                    
                \end{cases}
\end{equation}
$$WSCDS_{ij}(d) = \frac{(\textnormal{weight of model  i}) * s_{ij}(d)}{\textnormal{sum of weights}}$$
For WRCDS, replace $s_{ij}(d)$ with $r_{ij}(d)$ and $w_i$ with $\frac{1}{w_i}$. 

\section{Evaluation}

\begin{table*}[t]
\caption{Two model combination - Weighted score combinations by performance (recalls) for the data items $d_{1}$ - $d_{10}$.}
\label{tab:wgt_score_comb_by_perf_recall}
\centering
\begin{adjustbox}{width=\textwidth}
\begin{tabular}{|c|c|c|c|c|c|c|c|c|c|c|c|c|c|c|c|}
\hline
$D_i$ & AB & AC & AD & AE & AF & BC & BD & BE & BF & CD & CE & CF & DE & DF & EF \\ 
\hline
$d_{1}$ & 1 & 0.76258 & 1 & 1 & 1 & 0.77052 & 1 & 1 & 1 & 0.77067 & 0.77069 & 0.77077 & 1 & 1 & 1 \\
$d_{2}$ & 1 & 0.99976 & 1 & 1 & 1 & 0.99976 & 1 & 1 & 1 & 0.99976 & 0.99976 & 0.99976 & 1 & 1 & 1 \\
$d_{3}$ & 1 & 0.54627 & 1 & 1 & 1 & 0.56144 & 1 & 1 & 1 & 0.56172 & 0.56175 & 0.56191 & 1 & 1 & 1 \\
$d_{4}$ & 1 & 0.54597 & 1 & 1 & 1 & 0.56115 & 1 & 1 & 1 & 0.56143 & 0.56146 & 0.56162 & 1 & 1 & 1 \\
$d_{5}$ & 0.99996 & 0.55247 & 1 & 1 & 1 & 0.52705 & 0.99996 & 0.99996 & 0.99996 & 0.56771 & 0.56774 & 0.5679 & 1 & 1 & 1 \\
$d_{6}$ & 1 & 0.54641 & 1 & 1 & 1 & 0.56158 & 1 & 1 & 1 & 0.56186 & 0.5619 & 0.56205 & 1 & 1 & 1 \\
$d_{7}$ & 1 & 0.98708 & 1 & 1 & 1 & 0.98635 & 1 & 1 & 1 & 0.98752 & 0.98752 & 0.98753 & 1 & 1 & 1 \\
$d_{8}$ & 1 & 0.98739 & 1 & 1 & 1 & 0.98727 & 1 & 1 & 1 & 0.98741 & 0.98744 & 0.98744 & 1 & 1 & 1 \\
$d_{9}$ & 1 & 0.53821 & 1 & 1 & 1 & 0.53929 & 1 & 1 & 1 & 0.53991 & 0.53991 & 0.53991 & 1 & 1 & 1 \\
$d_{10}$ & 0.76986 & 0.76639 & 0.78936 & 0.78938 & 0.78945 & 0.98505 & 0.99962 & 0.99962 & 0.99962 & 0.99845 & 0.99845 & 0.99845 & 1 & 1 & 1 \\
\hline
\end{tabular}
\end{adjustbox}
\end{table*}

\begin{table*}[t]
\caption{Rankings of the WSCP results presented in Table \ref{tab:wgt_score_comb_by_perf_recall} for the data items $d_{1}$ - $d_{10}$.}
\label{tab:rankings_wgt_comb_by_perf_recall}
\centering
\begin{adjustbox}{width=\textwidth}
\begin{tabular}{|c|c|c|c|c|c|c|c|c|c|c|c|c|c|c|c|}
\hline
$D_i$ & AB & AC & AD & AE & AF & BC & BD & BE & BF & CD & CE & CF & DE & DF & EF \\ 
\hline
$d_{1}$ & 1 & 114936 & 1 & 1 & 1 & 105351 & 1 & 1 & 1 & 115227 & 116010 & 115949 & 1 & 1 & 1 \\
$d_{2}$ & 1 & 16485 & 1 & 1 & 1 & 13230 & 1 & 1 & 1 & 17815 & 17547 & 17270 & 1 & 1 & 1 \\
$d_{3}$ & 1 & 176670 & 1 & 1 & 1 & 157873 & 1 & 1 & 1 & 175125 & 175153 & 175236 & 1 & 1 & 1 \\
$d_{4}$ & 1 & 179070 & 1 & 1 & 1 & 159554 & 1 & 1 & 1 & 177513 & 177561 & 177623 & 1 & 1 & 1 \\
$d_{5}$ & 176598 & 152728 & 1 & 1 & 1 & 207139 & 184970 & 185096 & 184279 & 152473 & 152519 & 15495 & 1 & 1 & 1 \\
$d_{6}$ & 1 & 175530 & 1 & 1 & 1 & 157132 & 1 & 1 & 1 & 174123 & 174147 & 174181 & 1 & 1 & 1 \\
$d_{7}$ & 1 & 37075 & 1 & 1 & 1 & 37065 & 1 & 1 & 1 & 38823 & 39053 & 38739 & 1 & 1 & 1 \\
$d_{8}$ & 1 & 36857 & 1 & 1 & 1 & 36161 & 1 & 1 & 1 & 38925 & 39134 & 38825 & 1 & 1 & 1 \\
$d_{9}$ & 1 & 197884 & 1 & 1 & 1 & 181049 & 1 & 1 & 1 & 198636 & 198655 & 198640 & 1 & 1 & 1 \\
$d_{10}$ & 219277 & 110590 & 218958 & 218912 & 218977 & 38342 & 199530 & 199659 & 198672 & 39536 & 39833 & 39516 & 1 & 1 & 1 \\
\hline
\end{tabular}
\end{adjustbox}
\end{table*}

\begin{table*}[t]
\centering
\caption{Performance evaluation of individual models at the class level of granularity.}
\begin{adjustbox}{width=\textwidth}
\label{tab:model_eval_fusion_tblOne}
\begin{tabular}{|ccc|ccc|ccc|ccc|ccc|ccc|ccc|}
\hline
&&& \multicolumn{3}{c|}{A} & \multicolumn{3}{c|}{B} & \multicolumn{3}{c|}{C} & \multicolumn{3}{c|}{D} & \multicolumn{3}{c|}{E} & \multicolumn{3}{c|}{F} \\
\hline
\textbf{$class$} &&&  Precision & Recall & F1score & Precision & Recall & F1score& Precision & Recall & F1score & Precision & Recall & F1score & Precision & Recall & F1score & Precision & Recall & F1score \\
\hline
$0$ &&& 0.995305 & 0.933387 & 0.963352        &   0.994616 & 0.169366 & 0.289444 & 0.9739 & 0.913225 & 0.942587  & 0.999228 & 0.998502 & 0.998865  & 0.996296 & 0.998638 & 0.997466 & 0.999691 & 0.999328 & 0.99951 \\
$1$ &&&  0.995305 & 0.933387 & 0.963352  &  0.096706 & 0.994536 & 0.176271 & 0 & 0 & 0 & 0.983696 & 0.989071 & 0.986376 & 0.994536 & 0.994536 & 0.994536  & 1 & 1 & 1 \\
$2$ &&&  0.042553 & 0.065574 & 0.051613  &   0.983337 & 0.999164 & 0.991187 & 0.783516 & 0.737625 & 0.759879 & 0.992537 & 0.978554 & 0.985496  & 0.999958 & 1 & 0.999979 & 1 & 1 & 1 \\
$3$ &&&  0.999791 & 0.999833 & 0.999812  &   0.356379 & 0.966292 & 0.520714 & 0.589041 & 0.81372 & 0.683387 & 0.960603 & 0.980485 & 0.970442 & 0.968347 & 0.976937 & 0.972623  & 0.969873 & 0.989947 & 0.979807 \\
$4$ &&&  0.778788 & 0.911886 & 0.840098  &   0.98708 & 0.974564 & 0.980782  & 0.723168 & 0.862027 & 0.786516 & 0.986929 & 0.995396 & 0.991144  & 0.999874 & 0.999723 & 0.999799 & 1 & 0.999824 & 0.999912 \\
$5$ &&&  0.811881 & 0.809211 & 0.810544  &   0.099778 & 0.814967 & 0.17779  & 0.684392 & 0.814967 & 0.743994 & 0.966503 & 0.972862 & 0.969672 & 0.9801 & 0.972039 & 0.976053 & 0.981224 & 0.988487 & 0.984842  \\
$6$ &&&  0.816667 & 0.794781 & 0.805575  &   0.034563 & 0.834274 & 0.066377 & 0.375488 & 0.33921 & 0.356428 & 0.975456 & 0.980959 & 0.9782 & 0.991507 & 0.988011 & 0.989756 & 0.992928 & 0.990127 & 0.991525 \\
$7$ &&&  0.688797 & 0.996 & 0.814391     &   0.996004 & 0.997 & 0.996502  & 0.531576 & 0.968 & 0.686281 & 0.996004 & 0.997 & 0.996502 & 0.999001 & 1 & 0.9995 & 1 & 1 & 1 \\
$8$ &&&  0.333333 & 1 & 0.5              &   1 & 1 & 1  & 0 & 0 & 0 & 1 & 0.5 & 0.666667 & 1 & 1 & 1  & 1 & 1 & 1 \\
$9$ &&&  0.958722 & 0.9926 & 0.975367    &   0.471534 & 0.997257 & 0.64031  & 0.99932 & 0.998339 & 0.998829 & 0.999321 & 0.999799 & 0.99956 & 0.999648 & 0.999924 & 0.999786 & 0.999648 & 0.99995 & 0.999799 \\
$10$ &&& 0.541325 & 0.983762 & 0.698367  &   0.941482 & 0.979702 & 0.960212 & 0.851852 & 0.622463 & 0.719312 & 0.987887 & 0.993234 & 0.990553 & 0.99594 & 0.99594 & 0.99594 & 0.998645 & 0.997294 & 0.997969 \\
$11$ &&& 0.017903 & 0.061765 & 0.027759  &   0.037879 & 0.102941 & 0.05538  & 0 & 0 & 0  & 0.706941 & 0.808824 & 0.754458 & 0.594937 & 0.414706 & 0.488735 & 0.715013 & 0.826471 & 0.766712 \\
$12$ &&& 0.001724 & 0.333333 & 0.003431  &   0.04918 & 1 & 0.09375 & 0 & 0 & 0   & 0 & 0 & 0  & 0 & 0 & 0 & 0 & 0 & 0 \\
$13$ &&& 0.052947 & 0.97546 & 0.100442   &   0.274874 & 1 & 0.431217 & 0 & 0 & 0  & 0.452991 & 0.325153 & 0.378571  & 1 & 0.018405 & 0.036145 & 0.46789 & 0.312883 & 0.375 \\
\hline
\end{tabular}
\end{adjustbox}
\end{table*}

\begin{figure}[!ht]
	\centering
    \includegraphics[height=5cm, width=8.8cm]{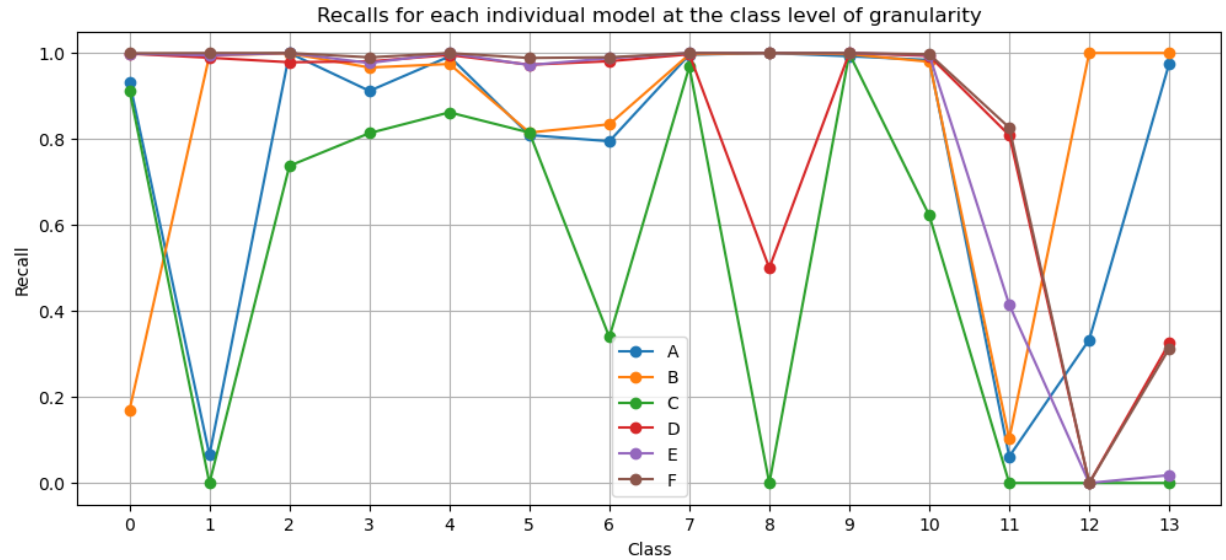}
    \caption{Performance of models at the class level of granularity.}
	\label{fig:plot_of_model_recalls_combination}
\end{figure}

\begin{figure*}[!ht]
	\centering
	\includegraphics[height=6.6cm, width=14cm]{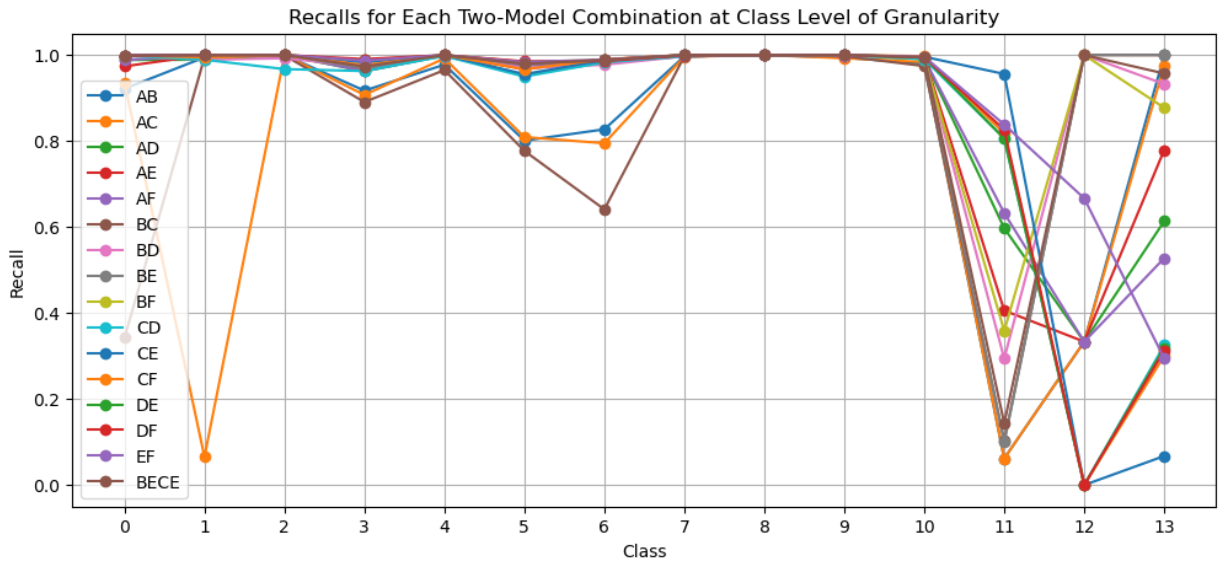}
    \caption{Recalls of each of the 15 combined models using weighted score combination by performance at the class level of granularity.
    }
	\label{fig:plot_of_model_recalls}
\end{figure*}



We assess model performance and efficiency at the class level, particularly in smaller sampled classes. Precise detection and classification of every attack are crucial, as even one missed attack can be harmful. Thus, our analysis emphasizes recall as a vital criterion. We employ diverse CFA metrics to combine models, harnessing strengths and addressing weaknesses. We calculate average score combination (ASC) and average rank combination (ARC) using the top 10 flows from 220,312, selected based on ASC and ARC metrics. These flows show high confidence in the fusion model's predictions. A small subset of the CFA dataset has all models with the highest probability scores, indicating lack of diversity. This tie results in identical top-ranking positions for both metrics. 
Although they yield the same outcome at rank 1, differences arise as rankings progress due to score variations. These metrics demonstrate distinct behavior with diverse scoring systems.


The weighted score combination by diversity strength and weighted rank combination by diversity strength are calculated, where the flows represent the first 10 out of a total of 4187 top-ranked flows. These flows are the ones in which the fusion model exhibits the highest confidence level in its prediction. Due to a lack of model diversity at the top-performing data items, the two above metrics produced the same output at rank position 1. However, the results vary as we move down the ranks. Table \ref{tab:wgt_score_comb_by_perf_recall} shows the two-model combination, weighted score combinations by performance (recalls) for the data items $d_{1}$ - $d_{10}$. Recall focuses on the model's ability to find all positive instances, measured per each attack class for all models. Last, Table \ref{tab:rankings_wgt_comb_by_perf_recall} summarizes the rankings of the WSCP results presented in the tables mentioned above for the data items $d_{1}$ - $d_{10}$. Building upon these tables, we aim to identify a model that performs well not only on the commonly observed attacks but also on the low-profiled ones.



While most models individually performed well against many attacks, they faced difficulty with low-profile attacks. Conversely, models excelling at low-profile attacks struggled with other attack types. This prompted us to consider model fusion, leveraging strengths and offsetting weaknesses. Our goal was to create a robust, comprehensive model by combining individual outputs, capable of effectively handling low-profile attacks and other traffic. Our method modifies the soft Voting Classifier as a metric for generating combined predictions. We average probability scores across classes for each model, using model recall values as weights for accurate aggregation. This advanced CFA technique, called weighted combination by performance, incorporates each model's recall performance at the attack level, enhancing fusion with model-specific attack data.

By assigning higher weights to models with higher recall rates for a given attack, we aimed to prioritize the models that were more adept at correctly identifying all attacks, including the low-profiled ones. This CFA strategy proved to be highly effective in enhancing the overall performance of our methodology. By incorporating the strengths of multiple models, we established a balanced and robust classification system that improved recall and reliability. Consequently, we successfully leveraged the individual strengths of each model, resulting in a more comprehensive and successful approach to attack identification and classification.

Table \ref{tab:model_eval_fusion_tblOne} highlights the performance of individual models, with Model $F$ emerging as the highest performer overall. However, it exhibited significant weaknesses in accurately classifying the low-profile attack class 12. On the other hand, Model $B$ demonstrated a flawless performance in identifying low-profiled attacks. Nonetheless, it fell short of effectively classifying other types of attacks compared to other models. Figures \ref{fig:plot_of_model_recalls_combination} and \ref{fig:plot_of_model_recalls} visually depict the performance of the six individual models and their two-model combinations across various classes, allowing for a comparison of their respective performances. A best performing models for each attack can be found in Table \ref{tab:best_performing_models}.


Among the various models, Model $BE$ consistently outperformed the others, exhibiting the highest recall rates for both low-profiled attacks and other traffic categories. This indicates that Model $BE$ excels in accurately identifying and classifying attacks, particularly in cases where limited training data is available. However, it should be noted that Model $BE$ obtained a recall of only 0.1 for attack class 11 which is a potential weakness. Model $CE$ demonstrated the highest recall of 0.9555882 for attack class 11, surpassing all other models. Motivated by this observation, we combined Models $BE$ and $CE$ to explore the possibility of achieving improved recalls across all attack classes. However, the fusion of these models, Model $BECE$, did not significantly increase the recall for Class 11. 

Table \ref{tab:best_performing_models} presents a summary of the performance for individual and combined models. Notably, Model $DF$ achieved a remarkable 100\% recall for the attack class 4, surpassing the performance of the best individual model in that category. This highlights the effectiveness of the combined model approach in achieving higher recall rates. Based on the CFA metric employed, our recommended fusion model would be to use Model $DF$ for attack class 4 and then Model $BE$ for all attack classes except class 11, for which Model $CE$ is employed. This combination would yield better overall performance in terms of recall and outperform the individual models. 

Last, while Model $BE$ and Model $CE$ successfully enhanced the recall rates for all attack classes, developing more advanced fusion techniques may yield even better results. Overall, our approach of combining models proved successful in achieving our goal of developing a model that performs well, particularly in recalling low-profiled attacks.

\begin{table}[h!]
\caption{Best performing models for each class (by recalls). Multiple models for a given class represents a tie.}
\label{tab:best_performing_models}
\centering
\begin{tabular}{|l|l|l|}
\hline
Traffic & Individual & Combined using WSCP \\
\hline
0 & F (99.93) & CF (99.92) \\ \hline
1 & F (100) & AE, AF, BE, CE, DE, DF, EF, BECE \\ & & (100) \\ \hline
2 & F (100) &BF, CE, CF, DE, BECE, EF (100 )\\ \hline
3 & A (99.98) & DF (99.05) \\\hline
4 & F (99.98) & DF (100) \\ \hline
5 & F (98.85) & DF (98.52) \\ \hline
6 & F (99.01) & CF (99.01) \\ \hline
7 & E, F (100) & AE, BE, CE, CF, DE, BECE, EF (100) \\ \hline
8 & A, B, E, F & AB, AC, AD, AE, AF, BC, EF, BD (100) \\ 
  &   &  BE, BF, CD, CE, CF, DE, DF, EF, BECE  \\ \hline
9 & F (100) & BF, CF (100) \\ \hline
10 & F (99.73) & CF (99.73) \\ \hline
11 & F (82.65) & CE (95.59) \\ \hline
12 & B (100) & BF, BECE, BC, BD, BE (100)\\ \hline
13 & B (100) & AB, BC, BE (100) \\ \hline
\end{tabular}
\end{table}

\label{sec:evaluation}

\section{Conclusion} This work uses the fusion approach to enhance ML-based DoS attack detection. Specifically, we targeted low-profile attack detection, an area where single models faltered. By merging multiple ML models with advanced CFA metrics, we improved precision, recall, and F1-score. This produced a highly effective combined model that nearly perfectly detected all attacks, including low-profile ones, achieving 100\% recall. Unlike most single models, our approach succeeded in this. CFA methods hold promise for boosting DoS attack detection. Future work can refine algorithms, explore more scoring systems, and integrate emerging tech for stronger defense. Advancing in this field ensures secure online services despite evolving threats.

 \label{sec:conclusion}


\bibliographystyle{IEEEtran}
\bibliography{refs}

\begin{thebibliography}{10}
\providecommand{\url}[1]{#1}
\csname url@samestyle\endcsname
\providecommand{\newblock}{\relax}
\providecommand{\bibinfo}[2]{#2}
\providecommand{\BIBentrySTDinterwordspacing}{\spaceskip=0pt\relax}
\providecommand{\BIBentryALTinterwordstretchfactor}{4}
\providecommand{\BIBentryALTinterwordspacing}{\spaceskip=\fontdimen2\font plus
\BIBentryALTinterwordstretchfactor\fontdimen3\font minus
  \fontdimen4\font\relax}
\providecommand{\BIBforeignlanguage}[2]{{%
\expandafter\ifx\csname l@#1\endcsname\relax
\typeout{** WARNING: IEEEtran.bst: No hyphenation pattern has been}%
\typeout{** loaded for the language `#1'. Using the pattern for}%
\typeout{** the default language instead.}%
\else
\language=\csname l@#1\endcsname
\fi
#2}}
\providecommand{\BIBdecl}{\relax}
\BIBdecl

\bibitem{gupta2021distributed}
B.~B. Gupta and A.~Dahiya, \emph{Distributed Denial of Service (DDoS) Attacks:
  Classification, Attacks, Challenges and Countermeasures}.\hskip 1em plus
  0.5em minus 0.4em\relax CRC press, 2021.

\bibitem{rahouti2021synguard}
M.~Rahouti, K.~Xiong, N.~Ghani, and F.~Shaikh, ``\mbox{SYNGuard}: Dynamic
  threshold-based \mbox{SYN} flood attack detection and mitigation in
  software-defined networks,'' \emph{IET Networks}, vol.~10, no.~2, pp. 76--87,
  2021.

\bibitem{mittal2022deep}
M.~Mittal, K.~Kumar, and S.~Behal, ``Deep learning approaches for detecting
  \mbox{DDoS} attacks: A systematic review,'' \emph{Soft Computing}, pp. 1--37,
  2022.

\bibitem{zhijun2020low}
W.~Zhijun, L.~Wenjing, L.~Liang, and Y.~Meng, ``Low-rate \mbox{DoS} attacks,
  detection, defense, and challenges: a survey,'' \emph{IEEE Access}, vol.~8,
  pp. 43\,920--43\,943, 2020.

\bibitem{david2021discriminating}
J.~David and C.~Thomas, ``Discriminating flash crowds from \mbox{DDoS} attacks
  using efficient thresholding algorithm,'' \emph{JPDC}, vol. 152, pp. 79--87,
  2021, Elsevier.

\bibitem{ali2023machine}
T.~E. Ali, Y.-W. Chong, and S.~Manickam, ``Machine learning techniques to
  detect a \mbox{DDoS} attack in \mbox{SDN}: A systematic review,''
  \emph{Applied Sciences}, vol.~13, no.~5, p. 3183, 2023.

\bibitem{deepa2019design}
V.~Deepa, K.~M. Sudar, and P.~Deepalakshmi, ``Design of ensemble learning
  methods for \mbox{DDoS} detection in \mbox{SDN} environment,'' in
  \emph{ViTECoN}.\hskip 1em plus 0.5em minus 0.4em\relax IEEE, 2019, pp. 1--6.

\bibitem{hsu2006combinatorial}
D.~F. Hsu, Y.-S. Chung, and B.~S. Kristal, ``Combinatorial fusion analysis:
  methods and practices of combining multiple scoring systems,'' in
  \emph{Advanced data mining technologies in bioinformatics}.\hskip 1em plus
  0.5em minus 0.4em\relax IGI Global, 2006, pp. 32--62.

\bibitem{hsu2010rank}
D.~F. Hsu, B.~S. Kristal, and C.~Schweikert, ``Rank-score characteristics
  (\mbox{RSC}) function and cognitive diversity,'' in \emph{International
  Conference on Brain Informatics}.\hskip 1em plus 0.5em minus 0.4em\relax
  Springer, 2010, pp. 42--54.

\bibitem{hurley2020multi}
L.~Hurley, B.~S. Kristal, S.~Sirimulla, C.~Schweikert, and D.~F. Hsu,
  ``Multi-layer combinatorial fusion using cognitive diversity,'' \emph{IEEE
  Access}, vol.~9, pp. 3919--3935, 2020.

\bibitem{tian2021lightweight}
Q.~Tian, C.~Guang, C.~Wenchao, and W.~Si, ``A lightweight residual networks
  framework for \mbox{DDoS} attack classification based on federated
  learning,'' in \emph{INFOCOM WKSHPS}.\hskip 1em plus 0.5em minus 0.4em\relax
  IEEE, 2021, pp. 1--6.

\bibitem{rosay2021cic}
\BIBentryALTinterwordspacing
A.~Rosay, F.~Carlier, E.~Cheval, and P.~Leroux, ``From {CIC-IDS2017} to
  {LYCOS-IDS2017}: A corrected dataset for better performance,'' in
  \emph{IEEE/WIC/ACM WI-IAT}.\hskip 1em plus 0.5em minus 0.4em\relax ACM, 2021,
  p.~6. [Online]. Available: \url{https://doi.org/10.1145/3486622.3493973}
\BIBentrySTDinterwordspacing

\bibitem{hsu2002methods}
D.~Hsu, J.~Shapiro, and I.~Taksa, ``Methods of data fusion in information
  retreival: Rank vs. score combination,'' DIMACS TR 2002, Tech. Rep., 2002.

\bibitem{hsu2019cognitive}
D.~F. Hsu, B.~S. Kristal, Y.~Hao, and C.~Schweikert, ``Cognitive diversity: A
  measurement of dissimilarity between multiple scoring systems,''
  \emph{Journal of Interconnection Networks}, vol.~19, no.~01, p. 1940001,
  2019.

\end{thebibliography}

\end{document}